\documentclass[11pt,a4paper]{article}
\pdfoutput=1

\pdfoutput=1
\usepackage{jcappub}
\usepackage[utf8]{inputenc}
\usepackage{graphicx}
\usepackage{color}
\usepackage{xfrac}
\usepackage{verbatim}
 \usepackage{mathtools}
\newcommand{\bea}{\begin{equation}\begin{aligned}}
\newcommand{\eea}{\end{aligned}\end{equation}}
\def\lsim{\mathrel{\raise.3ex\hbox{$<$\kern-.75em\lower1ex\hbox{$\sim$}}}}
\def\gsim{\mathrel{\raise.3ex\hbox{$>$\kern-.75em\lower1ex\hbox{$\sim$}}}}

\providecommand{\f}[2]{\frac{{#1}}{{#2}}}

\newcommand{\ee}[1]{\begin{equation}#1\end{equation}}
\newcommand{\ea}[1]{\begin{align}#1\end{align}}

\renewcommand{\lesssim}{\mbox{\raisebox{-.9ex}{~$\stackrel{\mbox{$<$}}{\sim}$~}}}
\renewcommand{\gtrsim}{\mbox{\raisebox{-.9ex}{~$\stackrel{\mbox{$>$}}{\sim}$~}}}

\title{Non-minimal gravitational reheating %and dark matter generation
  during kination}

\author[a]{Konstantinos Dimopoulos}
\author[b]{and Tommi Markkanen}
\affiliation[a]{Consortium for Fundamental Physics, Physics Department,
Lancaster University, Lancaster LA1 4YB, United Kingdom}

\affiliation[b]{Department of Physics, Imperial College London,\\Blackett Laboratory, London, SW7 2AZ, United Kingdom}
                            
\emailAdd{konst.dimopoulos@lancaster.ac.uk}                            
\emailAdd{t.markkanen@imperial.ac.uk}

\abstract{
  A new mechanism is presented which can reheat the Universe in non-oscillatory
  models of inflation, where the inflation period is followed by a period
  dominated by the kinetic density for the inflaton field (kination). The
  mechanism considers an auxiliary field non-minimally coupled to gravity. The
  auxiliary field is a spectator during inflation, rendered heavy by the
  non-minimal coupling to gravity. During kination however, the non-minimal
  coupling generates a tachyonic mass, which displaces the field, until its bare
  mass becomes important, leading to coherent oscillations. Then, the field
  decays into the radiation bath of the hot big bang. The model is
  generic and predictive, in that the resulting reheating temperature is a
  function only of the model parameters (masses and couplings) and not of
  initial conditions. It is shown that reheating can be very efficient also
  when considering only the Standard Model.
}

\begin{document}
\begin{flushleft}
%	\hfill		  LANCASTER/XXX\\
	\hfill		  IMPERIAL/TP/2018/TM/01
\end{flushleft}
\maketitle

%%%%%%%%%%%%%%%%%%%%%%%%%%%%%%%%%%%
\section{Introduction}
The most compelling mechanism for the generation of the primordial curvature
perturbation which is responsible for the formation of structures in the
Universe is the theory of Cosmic Inflation, which postulates that, in the
beginning of its history, our Universe underwent a period of accelerated
expansion. Inflation also accounts for the substantial fine-tuning of the
initial conditions of Hot Big Bang cosmology, namely the infamous horizon and
flatness problems. Observations of ever increasing precision are in definite
support of the scenario of inflation. In particular, the observed primordial
anisotropy in the Cosmic Microwave Background (CMB) radiation, suggests that
the curvature perturbation is predominantly adiabatic and Gaussian and almost 
(but not quite) scale-invariant \cite{planck}, in agreement with the simplest 
and most generic inflationary predictions.

In particle cosmology, inflation is realised through the inflationary paradigm,
which considers that the Universe undergoes accelerated expansion when dominated
by the potential density of a homogeneous scalar field. After the observation of
the electroweak Higgs field, we are confident that fundamental scalar fields
exist. One such field, called the inflaton, is assumed to control the dynamics
of inflation. Recent CMB observations, from the Planck satellite, strongly
suggest that the inflaton's scalar potential features a plateau with density
roughly $10^{-10}M_{\rm P}^4$ \cite{planck,BICEP2}. When the expectation value of 
the field is such that corresponds to the inflationary plateau, its potential 
density is roughly constant as the field slowly varies (rolls) along the 
plateau. At some point, the inflaton reaches the end of the plateau, where the 
scalar potential becomes steep and curved in field space, such that the field 
ceases to be potentially dominated and inflation ends. Traditionally, the 
inflationary paradigm suggests that the inflaton's scalar potential features a 
potential minimum not very far (in field space) from the inflationary plateau, 
such that, after inflation ends, the inflaton field oscillates around its 
vacuum expectation value (VEV) before decaying (perturbatively or not) into the 
thermal bath of the Hot Big Bang. The process is called (p)reheating.

However, in contrast to plateau inflation, which is supported by observations,
the oscillatory reheating scenario described above is merely a conjecture.
Indeed, many inflation models with a scalar potential that does not feature a
minimum near the inflationary plateau have been considered over the years.
Typically, the VEV of the inflaton field is displaced to infinity, such that,
after the end of inflation, the field rolls down the steep runaway potential,
being dominated by its kinetic density. This type of inflation is called
non-oscillatory \cite{NO}. The period of kinetic density domination, which 
follows inflation, is called kination \cite{kination}. The attractive feature
of such models is 
that the inflaton field avoids decay, which means that it can survive until 
today and play the role of quintessence (leading to late-time inflation) 
\cite{Q}, which addresses the current dark energy observations that suggest that
the Universe at present undergoes accelerated expansion again. Such 
quintessential inflation models \cite{QI} employ a single degree of freedom (the
inflaton field) to provide a unified description of inflation and dark
energy in the context of a common theoretical framework. However,
non-oscillatory inflation can have other consequences as well, and does not
need to be associated with dark energy. For example, a period of kination
results in the number of inflationary e-folds corresponding to the cosmological
scales exceeding the usual 60, which may have profound implications on the
values of inflationary observables \cite{lindeQ}.

Since the inflaton field does not decay in non-oscillatory inflation models (as
it does not have a particle interpretation) the formation of the thermal bath
of the Hot Big Bang has to be achieved by other means. After reheating is
somehow managed, the roll of the inflaton is halted and the field freezes
at some value substantially displaced from the inflationary plateau, with
some non-zero potential density, which can be the dark energy at present 
\cite{QI}. The most minimal approach to reheating in non-oscillatory inflation,
which agrees with the economy of the quintessential inflation proposal, is the
so-called gravitational reheating \cite{gravreh}. This is due to particle 
production of all light and conformally non-invariant fields during inflation, 
which generates a thermal bath characterised by the Hawking temperature of de 
Sitter space \mbox{$T_{\rm H}=H/2\pi$}, where $H$ is the inflationary Hubble 
scale. Gravitational reheating is inescapable but very inefficient. This means 
that kination lasts for a long time, before the generated radiation reheats the
Universe and freezes the rolling field. Unfortunately, a prolonged kination
period results in a large spike in the tensor power spectrum at small scales
\cite{spike}. In gravitational reheating, this spike is large enough to 
challenge the process of Big Bang Nucleosynthesis (BBN), because the tensor 
density is almost 1\% of the total during BBN. Thus, a reheating mechanism more 
efficient that gravitational reheating is preferable.

A mechanism which has been employed to efficiently reheat the Universe during
kination is instant preheating \cite{instant}. This removes a fraction of the 
inflaton's kinetic density, through particle production when the inflaton 
crosses an Enhanced Symmetry Point (ESP) on the way down along its runaway 
potential, such that the mass of another field coupled to the inflaton changes 
non-adiabaticaly. This can be a very efficient mechanism, which easily overcomes
the problem of excessive tensors. However, it needs to presuppose the existence 
of a suitable ESP at the right place in the inflaton direction, which means that
it is rather restrictive model-building-wise. This is why a third option for 
reheating the Universe in non-oscillatory inflationary models has been 
considered, namely curvaton reheating \cite{curvreh}.

The curvaton reheating mechanism considers the presence of a suitable auxiliary
scalar field, called the curvaton \cite{curvaton}. The field is a spectator 
during inflation. It is assumed light such that it undergoes particle production
during inflation, and develops thereby stochasticaly an expectation value 
displaced from the minimum along the curvaton direction. After the end of 
inflation, because the field is originally light, it remains frozen until the 
expansion rate decreases enough for the field to become heavy and begin 
coherent oscillations around its VEV. The decay of the curvaton field generates 
the desired radiation bath of the Hot Big Bang. Now, in the original curvaton 
proposal \cite{curvaton}, the curvaton field also generated the dominant 
contribution to the curvature perturbations (hence it's name). For the reheating
purposes however, this is not necessary and the curvaton's contribution to the 
curvature perturbations can be negligible.%
\footnote{The curvaton may decay into the radiation bath before or after it 
dominates the Universe. It may even lead to a brief period of inflation
after it dominates but before it decays into radiation
\cite{Dimopoulos:2011gb}.} 
This proposal has the merit that no interaction is required with the inflaton,
which means that inflation model-building is liberated \cite{liber}. 
However, it suffers from some drawbacks. In particular, the requirement that 
the curvaton remains light during inflation is non-trivial and amounts to some 
fine-tuning, especially in supergravity theories, in which masses of order $H$ 
are expected for all scalar fields, including the inflaton ($\eta$-problem)
\cite{DRT}. Also, the expectation value of the curvaton during inflation is a 
crucial unknown parameter, which determines the generated reheating temperature,
but depends strongly on the total duration of inflation \cite{wands},
that is unknowable because of the no hair theorem. This strongly
undermines the predictability of the proposal. 

In this paper we propose a new mechanism which reheats the Universe in a
non-oscillatory inflation model. Our mechanism makes use of the tendency of
light scalar fields to become excited by the expansion of space, which happens
quite generically in most cosmological spacetimes \cite{Markkanen:2017edu}.
It enjoys the advantages of instant
preheating and curvaton reheating, in that it can be highly efficient
but also avoids their disadvantages in terms of not having an impact on
inflation model-building (as has instant preheating, which requires a suitable
ESP) and not requiring the appropriate tuning of initial conditions 
(as with curvaton reheating, which needs to specify the expectation value of the
curvaton field in inflation). We consider the presence of an auxiliary scalar
field, which is a spectator during inflation, much like the curvaton in curvaton
reheating. However, our field is non-minimally coupled to gravity, which is
a generic expectation as setting the non-minimal coupling to zero is not stable
with respect to radiative corrections \cite{Chernikov:1968zm}. For a related
reheating model making use of the non-minimal coupling, see
Ref.~\cite{Figueroa:2016dsc}.  During inflation this coupling makes the field
heavy and it settles in the potential minimum, in contrast 
to the curvaton proposal. After inflation and during kination the inflationary 
minimum becomes a maximum in the scalar potential of our auxiliary field, which 
displaces the field at a value depended only on its non-minimal coupling to 
gravity (and possibly its self-coupling). Eventually, the bare mass of our 
field becomes important and the field oscillates around its VEV. It then may 
decay into the thermal bath of the Hot Big Bang. Our mechanism is generic and 
predictive because the generated reheating temperature depends only on the 
value of our model parameters such as masses and couplings.
%
%In a variant of our proposal, in which our field does not
%decay or it decays into heavy fermions, we also show that we can produce the
%dark matter in the Universe instead of the radiation bath, provided reheating 
%is achieved otherwise.

We consider natural units, where \mbox{$c=\hbar=k_B=1$} and
$M_{\rm P}^2\equiv (8\pi G)^{-1}$, with $G$ being Newton's gravitational constant
and \mbox{$M_{\rm P}=2.43\times 10^{18}\,$GeV} being the reduced Planck mass.
Our conventions are (+,+,+) \cite{Misner:1974qy}.

\section{The model}

In this work we consider a non-oscillatory model of inflation in which, after
the end of inflation, the influence of the potential to the field dynamics
practically disappears leading to a phase of kinetic density domination
(kination), i.e. an epoch that is described with the cosmological equation
of state $w=p_\phi/\rho_\phi=1$, where $p_\phi$ and $\rho_\phi$ are the pressure
and energy density of the inflaton field respectively. Besides considering a
kination epoch after inflation we leave the specifics of inflation unfixed.
%(in particular treat the scale of inflation as a free parameter).

In addition to the inflaton we assume the existence of an originally
subdominant, non-minimally coupled and massive scalar field with quartic
self-interactions
\ee{S_\chi=-\int d^4x\,\sqrt{|g|}\bigg[\f{1}{2}(\partial_\mu\chi)^2+
    \f{1}{2}m^2\chi^2+\f{1}{2}\xi R\chi^2+\f{1}{4}\lambda\chi^4\bigg]\,,
  \label{eq:act}}
where $\xi\geq1$. Importantly, we assume inflation to occur at such a high
scale that during much of the kination period the bare mass $m$ of field is
negligible with respect to the background Hubble rate.

Furthermore, the $\chi$-sector is assumed, for all practical purposes, to be
decoupled from the inflaton. It is however assumed to have sizeable couplings to
the Standard Model (SM) degrees of freedom. Therefore, $\chi$ could for example
be the length of the Higgs doublet, but in order to keep our analysis as
general as possible, the $\chi$-field may also be a generic beyond-the-SM field,
with sufficient couplings to the SM. 

\section{Condensate creation during kination}
\label{sec:con}

The potential for the $\chi$-field can be read from (\ref{eq:act})
\ee{V(\chi)=\f{1}{2}m^2\chi^2+\f{1}{2}\xi R\chi^2+\f{1}{4}\lambda\chi^4\,,
  \label{eq:pot}}
where $R=3(1-3w)H^2$ is the curvature scalar.
As we mentioned in the introduction, after the inflationary epoch when $w=-1$,
the inflaton enters a period of kination with the equation of state $w=1$. By
using the Friedmann equations 
\ea{
  \begin{cases}\phantom{-(}3H^2M_{\rm P}^2&= \rho_\phi\\
    -(3H^2+2\dot{H})M_{\rm P}^2 &= p_\phi=w\rho_\phi \end{cases}\,,\label{eq:e}}
it is straightforward to investigate the scale factor, the Hubble rate and the
scalar curvature for a given equation of state parameter $w$.
We have
\begin{eqnarray}
{\rm Inflation:}\,\qquad
  a\propto e^{Ht}\,, & H={\rm constant}\,, &  R=12H^2\label{eq:rehaH0}\,,\\
{\rm Kination:}\qquad
  a\propto t^{{1}/{3}}\,, & H=1/3t\,, & R=-6H^2\label{eq:rehaH}\,.
\end{eqnarray}
%During inflation they can be written as
%\ee{a\propto e^{Ht}\,,\quad H={\rm constant}\,,\quad R=12H^2\label{eq:rehaH0}\,,}
%and for kination as
%\ee{a\propto t^{{1}/{3}}\,,\quad H=\f{1}{3t}\,, \quad R=-6H^2\label{eq:rehaH}\,,}
%respectively. 

\begin{figure}
	\begin{center}
		\includegraphics[width=0.55\textwidth]{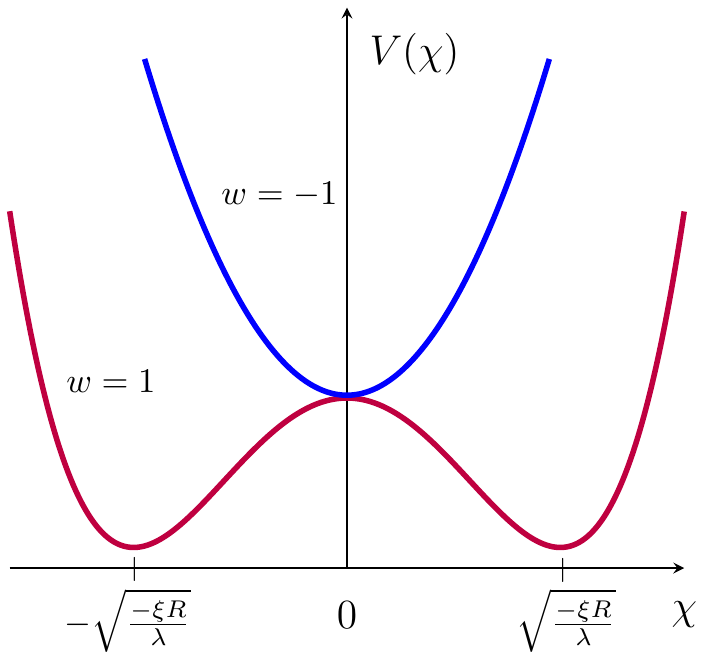}
	\end{center}
	\caption{The potential (\ref{eq:pot}) as a minimum at zero during inflation as described by the blue curve, but during kination if the tree-level mass parameter is smaller than $\sqrt{-\xi R^{\color{white}0}\!\!}$ the potential will posses a minimum displaced from zero, as depicted by the purple curve.\label{fig:sca}}
\end{figure}

\pagebreak

Since we are assuming a positive non-minimal coupling with $\xi\geq1$ and that
the Hubble rate during inflation is much larger than the bare mass of the field,
$H\gg m$, $\xi R$ gives the $\chi$-field a large positive mass parameter during
inflation but becomes negative during kination. Therefore, the vacuum state is
at zero during inflation
\ee{\label{eq:min}V'(\chi_{\rm min})=0\quad\Leftrightarrow\quad\chi_{\rm min}=0\,;
  \qquad w=-1\,,} but gets displaced from zero during kination as 
\ee{\label{eq:min0}V'(\chi_{\rm min})=0\quad\Leftrightarrow\quad \chi_{\rm min}^2 =
  -\f{\xi R+m^2}{\lambda}\approx-\f{\xi R}{\lambda}=\frac{6\xi}{\lambda}H^2\,;
  \qquad w=1\,,}
as depicted in Fig. \ref{fig:sca}. For now on, we refer to the non-zero value of
the $\chi$-field in the minimum as the condensate.%
\footnote{The phase transition, which results in the creation of the non-zero
condensate may also lead it the copious formation of topological defects. Such
defects disappear when the vacuum at zero is restored, but they might influence
the evolution of the Universe beforehand. The kind of defects produced, if any,
and their stability depends on the actual theory considered. For example, if the
$\chi$-field is complex then one may have the formation of cosmic strings,
whose network, however, soon reaches a scaling solution and does not affect the
Universe evolution. Similarly for higher order defects (textures). However,
domain walls or monopoles would indeed affect the evolution of the Universe.
Topological defects, even with transient existence, might lead to interesting
effects but we do not investigate this possibility here. Instead, we assume
that the evolution of the Universe is not affected by potential topological
defects, so we can ignore them.}

%Before kination during the inflationary period the effective mass parameter from the non-minimal coupling gives a large positive contribution for $\chi$, $R= 12H^2$, indicating that during inflation the minimum of the potential is at $\chi=0$ and there is no condensate generation. Furthermore, 

%As $R=-6H^2$ is a decreasing function during kination, in order to generate an adequate energy density for reheating the $\chi$-field grow to its final value $\chi_{\rm min}$ relatively soon after inflation has ended, which as we will show is indeed the case.

%\subsection{Growth of the zero mode after inflation}

We may investigate the behaviour of the $\chi$-field by making use of a
classical approximation i.e. study the behaviour of the zero mode. Since we are
assuming that inflation takes place at a scale for which $m\ll H$ and that
$\xi\geq1$ giving $\chi=0$ during inflation, the equation of motion at the
start of kination can be approximated by neglecting the mass and the quartic
term in Eq.~(\ref{eq:act})
\ee{\ddot{\chi}+3H\dot{\chi}+\xi R\chi=0\quad\Leftrightarrow
  \quad\ddot{\chi}+\f{1}{t}\dot{\chi}-\f{2\xi}{3t^2}\chi=0\,.}
After inflation, the $\chi$-field becomes tachyonic. Thus, it is ``kicked-off''
the origin (which becomes a potential hill) due to quantum fluctuations, which
are of typical amplitude $\delta\chi=$\newline
\mbox{$|m_{\rm eff}|/2\pi$} per Hubble time
$H^{-1}$, %\cite{tachyonic}
\cite{Markkanen:2017edu,Starobinsky:1994bd}, 
where $m_{\rm eff}$ is the tachyonic mass of the field, whose square is
\begin{equation}
m_{\rm eff}^2=V''=m^2+\xi R\approx -6\xi H^2,
\label{meff}
\end{equation}
where we ignored the bare mass of the $\chi$-field, which is \mbox{$m\ll H$}
near the end of inflation. Thus, as an initial condition we use
\mbox{$\chi_{\rm end}=\sqrt{6\xi}H_{\rm end}/2\pi$} and
\mbox{$\dot\chi_{\rm end}=H_{\rm end}\chi_{\rm end}$},
with `end' denoting the end of inflation.%
\footnote{The value of $\chi_{\rm end}$ would mildly change in the presence of
more $\chi$-fields. As shown in Ref.~\cite{RefPaper+}, $\chi_{\rm end}$ is
proportional to $\sqrt{\cal N}$, where $\cal N$ is the number of fields. This
would weakly affect our results but the effect is of order unity and can be
ignored, at least for a number of fields which is not too large.}

From the above we obtain
\begin{eqnarray}
\chi & = & \f{H_{\rm end}}{4\pi}\bigg[
    (\sqrt{6\xi}+1)\bigg(\f{t}{t_{\rm end}}\bigg)^{\sqrt{6\xi}/3} +
  (\sqrt{6\xi}-1)\bigg(\f{t}{t_{\rm end}}\bigg)^{-\sqrt{6\xi}/3}\bigg]\;\Rightarrow
\nonumber\\
%\chi(t\gg t_{\rm end})
\chi & \simeq & (\sqrt{6\xi}+1)
  \f{H_{\rm end}}{4\pi}\bigg[\bigg(\f{t}{t_{\rm end}}\bigg)^{{1}/3}\bigg]^{\sqrt{6\xi}}
  \propto a^{\sqrt{6\xi}}\quad{\rm when}\;t\gg t_{\rm end}\,.
\label{xroll}
\end{eqnarray}
Since $\chi_{\rm min}\propto a^{-3}$ the above indicates that the condensate
quickly reaches the value $\sim \chi_{\rm min}$. We can estimate the Hubble rate
when this occurs by comparing $\chi_{\rm min}$ in Eq.~(\ref{eq:min0}) to the
above giving the condition
\ee{
  %\f{\chi}{\chi_{\rm min}}\gtrsim1\qquad \Leftrightarrow\qquad
  \f{\sqrt\lambda}{2\pi}=
  \bigg(\f{H_m}{H_{\rm end}}\bigg)^{1+\sqrt{6\xi}/3}
  \label{eq:rehcond}\,,}
where the subscript `$m$' denotes the moment when the $\chi$-field reaches
the minimum of the potential \mbox{$\chi=\chi_{\rm min}$}.
%Due to the sharp power-law scaling in terms of $\xi$ the above will turn out not to constrain the parameter space in any significant manner.
Thus, we see that, considering a perturbative coupling $\lambda<4\pi^2$, a
regime always exists when the $\chi$-field is rolling down the central potential
hill but has not reached the minimum of the potential yet, i.e.
\mbox{$H_m<H_{\rm end}$} and \mbox{$\chi<\chi_{\rm min}$}.

\section{Reheating from the kination condensate}
\label{sec:reh}

For achieving successful reheating, the energy density of the $\chi$-field,
$\rho_\chi$ (or its decay products), must start to dominate the evolution of the
Universe over that of the inflaton $\rho_\phi$. It also must, before or after it
has become dominant, decay into light SM degrees of freedom that thermalize.
During kination we can see from Eqs.~(\ref{eq:e}) and (\ref{eq:rehaH}) that the
energy-density of the inflaton scales as $\rho_\phi\propto a^{-6}$ and since from
Eqs.~(\ref{eq:min}) and (\ref{eq:rehaH})
$\chi_{\rm min}^2\propto R\propto t^{-2}\propto a^{-6}$ we also
see that in order for the energy density $\rho_\chi$ to overtake $\rho_\phi$
simply generating the condensate is not enough, but it or the energy density
sourced by it must evolve to have a scaling less than $a^{-6}$. 

In this work for completeness we assume a specific reheating model where the
$\chi$-field is coupled to light fermions in the SM sector via a Yukawa-type
interaction $\sim g\bar{\psi}\psi\chi$ giving rise to the decay rate 
\ee{\Gamma = g^2\f{M}{8\pi}\,,\label{eq:dec}}
where $g$ is a dimensionless coupling constant and $M$ the effective mass
parameter for $\chi$. Note that this is not the tachyonic effective mass in
Eq.~\eqref{meff} but it is given by \mbox{$M=\sqrt{V''(\chi_{\rm min})}$}.
From the potential in Eq.~(\ref{eq:pot}), when $\sqrt{6\xi}H\gg m$ and $w=1$,
we can obtain the mass as follows
\begin{equation}
V''(\chi_{\rm min})=m^2+\xi R+3\lambda\chi_{\rm min}^2=
2(6\xi H^2-m^2)\simeq 12\xi H^2\,,
\label{M}
  \end{equation}
with \mbox{$\chi_{\rm min}=\sqrt{6\xi/\lambda}\;H$}. (cf. Eq.~\eqref{eq:min0}).
When $\sqrt{6\xi}H\ll m$ we have \mbox{$\chi_{\rm min}=0$} and so
\mbox{$M^2=m^2-6\xi H^2\simeq m^2$}, i.e. $M=m$. When \mbox{$\Gamma<H$},
the decay of $\chi$-particles into light fermions is exponentially suppressed.
This means that the energy density stored in the $\chi$-field decays into the
light fermions only when $\Gamma=H$. Actually, it is possible for the field
not to decay even if $\Gamma>H$ if \mbox{$\chi<\chi_{\rm min}$} and the field
is still rolling down the potential hill. This is because, only after reaching
the potential minimum can the field perform coherent oscillations and only an
oscillating scalar field has a particle interpretation and can decay into
fermions.%
\footnote{This decay of the scalar particles of the oscillating condensate
should not be confused with the process of particle production, which creates
particles due to the gravitational background during inflation. This process is
discussed briefly in Sec.~4.3. Gravitational production of fermions is indeed
possible \cite{RefPaper}, without the direct decay of the scalar condensate.}

In a model with a Yukawa interaction we can see that there are broadly two
distinct scenarios in which the decay may happen. Soon after inflation when the
kination condensate has been generated and we can still neglect the bare mass
parameter $m$, by using $M=\sqrt{-2\xi R}$ and Eq.~(\ref{eq:dec}) we obtain
\begin{equation}
\frac{\Gamma}{H}=g^2\frac{\sqrt{3\xi}}{4\pi}\,.
  \end{equation}
Thus, $\Gamma \geq H$ is possible when
\ee{g^2\geq\frac{4\pi}{\sqrt{3\xi}}\,.
%\ee{g^2\sqrt{\f{3\xi}{16\pi^2}}\geq1\,.
  \label{eq:dec0}}
When the above bound is satisfied the condensate practically decays immediately
when the field reaches the minimum of the potential \mbox{$\chi=\chi_{\rm min}$},
i.e. at $H_m$ (as it cannot decay before).
In contrast, when the above threshold is not satisfied, the $\chi$-field can
decay only after $R$ has decreased such that the bare mass dominates $M$,
the vacuum is restored at $\chi=0$  and one may approximate $M\approx m$
leading to a constant $\Gamma$, which eventually leads to the decay of the
condensate.

Based on the above, in this work we consider two main
scenarios for obtaining successful reheating as sourced by the condensate
generated during kination:
\begin{itemize}
\item[A)] The $\xi$ and $g$ do not satisfy the condition in Eq.~(\ref{eq:dec0}).
Thus, the condensate can decay only after the bare mass $m$ dominates over
$\sqrt{-\xi R}$ and the vacuum is restored at $\chi=0$. 
\item[B)] The condition in Eq.~(\ref{eq:dec0}) is satisfied and the
$\chi$-field decays into radiation immediately after reaching the minimum of the potential, which happens when \mbox{$H=H_m$}.
\end{itemize}
%The above two scenarios do not represent an exhaustive list of possible reheating dynamics but serve as useful representative cases as predicted by a Yukawa interaction.

%After $\rho_\chi$ has started to dominate is must thermalize. For completeness we assume that the $\chi$-field is coupled to two light fermions via a Yukawa-type interaction.  
\subsection{Case A: decay after vacuum restoration}
\label{sec:vr}
If the constraint in Eq.~(\ref{eq:dec0}) is not satisfied, the decay of the condensate does not take place near the end of inflation. After a sufficiently long time has passed, the scalar curvature decreases such that the bare mass term of the field can no longer be neglected. At this threshold, the minimum of the potential shifts back to $\chi=0$ restoring the vacuum configuration that was present prior to kination i.e. the blue curve in
Fig.~\ref{fig:sca}. Soon after this point, the non-minimal and quartic terms
become negligible and the field starts oscillating around $\chi=0$ in a
quadratic potential with the energy density
\ee{\rho_\chi\approx \f{1}{2}m^2\chi^2\,,}
which diluted as matter by the Universe expansion. We will approximate that
immediately after the threshold
\ee{m^2=-\xi R_{\rm vr}=6\xi H_{\rm vr}^2\,,\label{eq:equ}}
where `vr' stands for vacuum restoration, the density of the $\chi$-field starts
scaling as \mbox{$\rho_\chi\propto a^{-3}$}.
%with the initial condition 
Indeed, the requirement that \mbox{$H_{\rm vr}>\Gamma$}, using %that
\mbox{$\Gamma=g^2m/8\pi$}, results in the bound
\begin{equation}
g^2<\frac{8\pi}{\sqrt{6\xi}},
  \label{gbound0}
\end{equation}
which is (roughly) the reverse of the bound in Eq.~\eqref{eq:dec0}.

After reaching $H_{\rm vr}$ in Eq.~(\ref{eq:equ}), this scenario has two distinct
ways of proceeding: vacuum restoration (meaning domination by the bare mass)
can occur either before or after the field has reached the temporary minimum 
\mbox{$\chi_{\rm min}=\sqrt{-\xi R/\lambda}$}, given in Eq.~\eqref{eq:min0}.
In each case, the decay into radiation can occur either before or after the
condensate has become the dominant component of the Universe content.
%This mechanism is subject to the condition (\ref{eq:rehcond}) demanding a sufficiently long time period before (\ref{eq:equ}) is reached, which for $\xi\gtrsim1$ is quite easily satisfied.

\subsubsection{\boldmath
  Vacuum restoration after the field reaches $\chi_{\rm min}$}

From the requirement \mbox{$H_m>H_{\rm vr}$} it is straightforward to obtain
\begin{equation}
  \frac{\sqrt\lambda}{2\pi}>\bigg(\sqrt{6\xi}\;
  \frac{H_{\rm end}}{m}\bigg)^{-(1+\sqrt{6\xi}/3)}\,,
\label{lbound1}
  \end{equation}
where we used Eqs.~\eqref{eq:rehcond} and \eqref{eq:equ}.

First we address the situation when the condensate decays into radiation before
$\rho_\phi=\rho_\chi$. Using \mbox{$\rho_\phi\propto a^{-6}$}, the fact that the
density of radiation scales as \mbox{$\rho_r\propto a^{-4}$} and
Eq.~(\ref{eq:equ}) we find that radiation starts dominating the evolution when
%Hubble rate has reached
\ee{1\equiv\f{\rho_\phi}{\rho_r}\bigg\vert_{\rm reh}=
  \bigg(\f{a_{\rm vr}}{a_{\rm dec}}\bigg)^3
  \bigg(\f{a_{\rm dec}}{a_{\rm reh}}\bigg)^2 \f{\rho_\phi}{\rho_\chi}\bigg\vert_{\rm vr}
  =\bigg(\f{a_{\rm vr}}{a_{\rm dec}}\bigg)^3\bigg(\f{a_{\rm dec}}{a_{\rm reh}}\bigg)^2
  \f{\lambda M_{\rm P}^2}{\xi m^2}\,,\label{eq:dommi}}
where `dec' stands for decay and we considered that the condensate decays fully
\mbox{$\rho_\chi^{\rm dec}=\rho_r^{\rm dec}$}. With the help of Eqs.~(\ref{eq:rehaH})
and (\ref{eq:equ}), we straightforwardly get
\ee{\bigg(\f{a_{\rm vr}}{a_{\rm dec}}\bigg)^3=
  \f{\Gamma}{H_{\rm vr}}=\f{\sqrt{6\xi}\,\Gamma}{m}\,;\qquad
  \bigg(\f{a_{\rm dec}}{a_{\rm reh}}\bigg)^2=
  \bigg({\f{H_{\rm reh}}{\Gamma}}\bigg)^{2/3}\,,}
allowing us to write from Eq.~(\ref{eq:dommi})
\ee{H_{\rm reh}^{2/3}=
  \bigg(\f{\xi}{6\lambda^2\Gamma^{2/3}}\bigg)^{1/2}\f{m^3}{M_{\rm P}^2}\,;\qquad
  \Gamma\geq H_{\rm reh}\,.\label{eq:gam}}
The reheating temperature then comes straightforwardly using
\mbox{$3H_{\rm reh}^2M_{\rm P}^2=\f{\pi^2}{30}g_*T_{\rm reh}^4$}, which gives
\ee{%=\f{\xi}{2\lambda^2}\f{m^6}{M_{\rm P}^2}
%  \quad\Leftrightarrow\quad
  T_{\rm reh}=\bigg(\f{75}{2\pi^4}\bigg)^{1/8}\bigg(\f{\xi^{3/2}}{g_*\lambda^{3}}
  \bigg)^{1/4}\bigg(\f{m^9} {M_{\rm P}^4\Gamma}\bigg)^{1/4}
%T_{\rm reh}=\bigg(\f{15\xi}{\pi^2\lambda^2g^*}\bigg)^{1/4}\f{m^{3/2}}{\sqrt{M_{\rm P}}}
	\,,}
where $g_*$ is the number of effective relativistic degrees of freedom. Using
\mbox{$\Gamma=g^2m/8\pi$}, the above can be writen as
\begin{equation}
  T_{\rm reh}=\bigg(\frac{10}{3\pi g_*}\bigg)^{1/4}\frac{1}{\sqrt g}
  \bigg(\frac{\sqrt{6\xi}}{\lambda}\bigg)^{3/4}\frac{m^2}{M_{\rm P}}\,.
\label{Treh1}
\end{equation}
%In view of the bound in Eq.~\eqref{lbound1}, the above results in the bound
%\begin{equation}
%  T_{\rm reh}<\bigg(\frac{5\pi}{96g_*}\bigg)^{1/4}\frac{1}{\pi^2\sqrt g}
%  (\sqrt{6\xi})^{\frac12(\frac92+\sqrt{6\xi})}
%  \bigg(\frac{H_{\rm end}}{m}\bigg)^{\frac12(3+\sqrt{6\xi})}\frac{m^2}{M_{\rm P}}\,.
%  \label{Trehbound1}
%  \end{equation}

Now, let us consider the situation when the field decays after
$\rho_\chi=\rho_\phi$. In this case, we can easily write
\ee{\label{Trehtemp}
  T_{\rm reh}=\bigg(\f{90}{\pi^2g_*}M_{\rm P}^2 \Gamma^2\bigg)^{1/4}\,,}
where we used Eq.~(\ref{eq:gam}) at the limit $\Gamma= H_{\rm dom}$, with the
`dom' denoting the domination $\rho_\phi$ from $\rho_\chi$, i.e.
\mbox{$\rho_\phi^{\rm dom}=\rho_\chi^{\rm dom}$}. $H_{\rm dom}$ is straightforward to
calculate, as follows
\begin{equation}\hspace{-2cm}
  1\equiv\left.\frac{\rho_\phi}{\rho_\chi}\right|_{\rm dom}\!\!\!\!=
  \left.\frac{\rho_\phi}{\rho_\chi}\right|_{\rm vr}
  \bigg(\frac{a_{\rm vr}}{a_{\rm dom}}\bigg)^3\!\!=\frac{\lambda}{\xi}
  \bigg(\frac{M_{\rm P}}{m}\bigg)^2\frac{H_{\rm dom}}{H_{\rm vr}}
\;\Rightarrow\;H_{\rm dom}=\frac{\sqrt{6\xi}}{6\lambda}\frac{m^3}{M_{\rm P}^2}\,,
\hspace{-1cm}
\end{equation}
where we used \mbox{$\rho_\phi^{\rm vr}=
  3H_{\rm vr}^2M_{\rm P}^2=\frac{3}{6\xi}(mM_{\rm P})^2$} and 
\mbox{$\rho_\chi^{\rm vr}=\frac12 m^2\chi_{\rm vr}^2=m^4/2\lambda$}, with
\mbox{$H_{\rm vr}=m/\sqrt{6\xi}$} (cf. Eq.~\eqref{eq:equ}) and %also
\mbox{$\chi_{\rm vr}=\sqrt{\frac{6\xi}{\lambda}}H_{\rm vr}=m/\sqrt\lambda$}
(cf. Eq.~\eqref{eq:min0}). The requirement that \mbox{$\Gamma<H_{\rm dom}$}
results in
\begin{equation}
\Gamma<\bigg(\f{\xi}{6\lambda^2}\bigg)^{1/2}\f{m^3}{M_{\rm P}^2}
\quad\Leftrightarrow\quad
g<\sqrt{\frac{4\pi}{3\lambda}}(\sqrt{6\xi})^{1/2}\frac{m}{M_{\rm P}}\,, 
\label{gbound1}
\end{equation}
where we used that \mbox{$\Gamma=g^2 m/8\pi$}. Using this also in
Eq.~\eqref{Trehtemp}, we find
\begin{equation}
  T_{\rm reh}=\bigg(\frac{45}{32g_*}\bigg)^{1/4}\frac{g}{\pi}\sqrt{mM_{\rm P}}\,.
  \label{Treh2}
\end{equation}
The bounds in Eqs.~\eqref{lbound1} and \eqref{gbound1} reinforce each other.
Combining them together we get
\begin{equation}
  g<\frac{1}{\sqrt{3\pi}}(\sqrt{6\xi})^{\frac32+\sqrt{6\xi}/3}
  \bigg(\frac{H_{\rm end}}{m}\bigg)^{1+\sqrt{6\xi}/3}\frac{m}{M_{\rm P}}\,.
  \label{gbound}
  \end{equation}
%When the above is applied to Eq.~\eqref{Treh2} we obtain
%\begin{equation}
%  T_{\rm reh}<\bigg(\frac{5\pi^2}{32g_*}\bigg)^{1/4}\frac{1}{\pi^2}
%  (\sqrt{6\xi})^{\frac32+\frac13\sqrt{6\xi}}
%  \bigg(\frac{H_{\rm end}}{m}\bigg)^{1+\sqrt{6\xi}/3}
%  \bigg(\frac{m}{M_{\rm P}}\bigg)^{1/2}m\,.
%\label{Trehbound2}
%\end{equation}
%
%%Case A is illustrated in Fig. \ref{fig:scal} as the curve that starts out as blue.

\begin{figure}
	\begin{center}
		\includegraphics[width=0.55\textwidth]{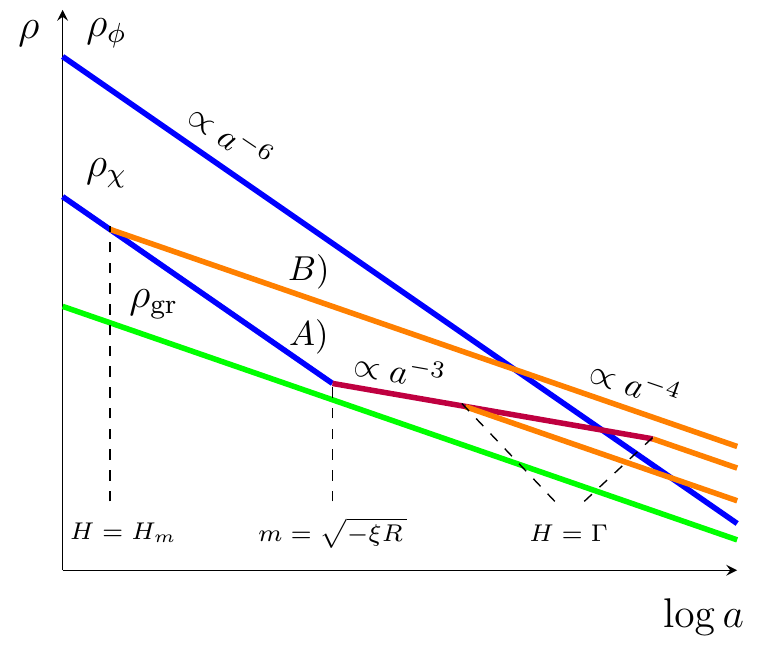}
	\end{center}
	\caption{Illustration of two possible reheating mechanisms sourced by the condensate generated during kination, where the decay into SM particles is assumed to proceed through a Yukawa-coupling. The top blue curve labelled $\rho_\phi$ represents the dominant component of the energy density of the Universe i.e. the inflaton sector. The blue curve labelled $\rho_\chi$ is the energy-density of the $\chi$ sector, which can either A) decay into radiation $\propto a^{-4}$ (orange lines) after vacuum restoration containing a period of dust-like evolution $\propto a^{-3}$ (the red line) or B) decay before this happens. The green curve represents the energy density produced by gravitational reheating (see Sec.~\ref{sec:gravr}).\label{fig:scal}}
\end{figure}

\subsubsection{\boldmath
  Vacuum restoration before the field reaches $\chi_{\rm min}$}

From the requirement \mbox{$H_m<H_{\rm vr}$} we obtain the reverse bound to the
one in Eq.~\eqref{lbound1}:
\begin{equation}
  \frac{\sqrt\lambda}{2\pi}<\bigg(\sqrt{6\xi}\;
  \frac{H_{\rm end}}{m}\bigg)^{-(1+\sqrt{6\xi}/3)}\,,
\label{lbound2}
  \end{equation}
where we used Eqs.~\eqref{eq:rehcond} and \eqref{eq:equ}.

At vacuum restoration the value of the condensate is obtained as follows.
Remember that, while the field is still rolling down the central potential
hill, we have \mbox{$\chi\propto a^{\sqrt{6\xi}}$} (cf. Eq.~\eqref{xroll}).
This means that
\begin{equation}
\chi_{\rm vr}=\chi_{\rm end}\bigg(\frac{a_{\rm vr}}{a_{\rm end}}\bigg)^{\sqrt{6\xi}}=
\frac{\sqrt{6\xi}H_{\rm end}}{2\pi}
\bigg(\frac{H_{\rm end}}{H_{\rm vr}}\bigg)^{\sqrt{6\xi}/3}\,,
  \end{equation}
where \mbox{$a\propto H^{-1/3}$} during kination.
Using this, we obtain
\begin{equation}
  \left.\frac{\rho_\phi}{\rho_\chi}\right|_{\rm vr}=
  \frac{3H_{\rm vr}^2M_{\rm P}^2}{\frac12m^2\chi_{\rm vr}^2}=
  24\pi^2(\sqrt{6\xi})^{-4-\frac23\sqrt{6\xi}}
  \bigg(\frac{m}{H_{\rm end}}\bigg)^{2(1+\sqrt{6\xi}/3)}
  \bigg(\frac{M_{\rm P}}{m}\bigg)^2\,,
\label{vrratio}
\end{equation}
where we used that \mbox{$H_{\rm vr}=m/\sqrt{6\xi}$} (cf. Eq.~\eqref{eq:equ}).

Consider first that the condensate decays after it dominates the Universe,
meaning \mbox{$\Gamma<H_{\rm dom}$}. $H_{\rm dom}$ can be easily obtained as
follows. 
\begin{equation}
1\equiv\left.\frac{\rho_\phi}{\rho_\chi}\right|_{\rm dom}=
\left.\frac{\rho_\phi}{\rho_\chi}\right|_{\rm vr}
\bigg(\frac{a_{\rm vr}}{a_{\rm dom}}\bigg)^3=
\left.\frac{\rho_\phi}{\rho_\chi}\right|_{\rm vr}\frac{H_{\rm dom}}{H_{\rm vr}}\,,
  \end{equation}
where we used \mbox{$\rho_\phi/\rho_\chi\propto a^{-3}$} and
\mbox{$a\propto H^{-1/3}$} during kination. Thus, considering also
Eq.~\eqref{eq:equ}, we get
\begin{equation}
H_{\rm dom}=\frac{1}{24\pi^2}(\sqrt{6\xi})^{3+\frac23\sqrt{6\xi}}
\bigg(\frac{H_{\rm end}}{m}\bigg)^{2(1+\sqrt{6\xi}/3)}\frac{m^3}{M_{\rm P}^2}\,.
\end{equation}
Demanding \mbox{$\Gamma<H_{\rm dom}$} results in a bound on $g$ identical to
Eq.~\eqref{gbound}. 

The reheating temperature is again given by Eq.~\eqref{Trehtemp} or
Eq.~\eqref{Treh2}.
%Applying the bound in Eq.~\eqref{gbound} we obtain again the
%bound in Eq.~\eqref{Trehbound2}.
Thus, we see that if reheating of the Universe is achieved by the decay of the
condensate into radiation after it has become the energetically dominant
component, it makes no difference whether or not it has reached the
potential minimum $\chi_{\rm min}$ before vacuum restoration.

Now, let us consider the case when the condensate decays before domination.
We find the reheating temperature as follows.
\begin{equation}
  1\equiv\left.\frac{\rho_\phi}{\rho_r}\right|_{\rm reh}=
\left.\frac{\rho_\phi}{\rho_\chi}\right|_{\rm vr}
\bigg(\frac{a_{\rm vr}}{a_{\rm dec}}\bigg)^3
\bigg(\frac{a_{\rm dec}}{a_{\rm reh}}\bigg)^2=
\left.\frac{\rho_\phi}{\rho_\chi}\right|_{\rm vr}
\frac{H_{\rm dec}}{H_{\rm vr}}\bigg(\frac{H_{\rm reh}}{H_{\rm dec}}\bigg)^{2/3}\,,
\end{equation}
where \mbox{$\rho_\chi^{\rm dec}=\rho_r^{\rm dec}$} and we considered that
\mbox{$\rho_\phi/\rho_r\propto a^{-2}$} and
\mbox{$\rho_\phi/\rho_\chi\propto a^{-3}$}, with \mbox{$a\propto H^{-1/3}$} during 
kination. Using Eq.~\eqref{vrratio} and considering that
\mbox{$H_{\rm vr}=m/\sqrt{6\xi}$} (cf. Eq.~\eqref{eq:equ}) and
\mbox{$H_{\rm dec}=\Gamma=g^2m/8\pi$}, we calculate $H_{\rm reh}$ and
subsequently obtain the reheating temperature, because
\mbox{$\frac{\pi^2}{30}g_*T_{\rm reh}^4=3H_{\rm reh}^2M_{\rm P}^2$}.
Thus, we find
\begin{equation}
  T_{\rm reh}=\bigg(\frac{5\pi}{96g_*}\bigg)^{1/4}\frac{1}{\pi^2\sqrt g}
  (\sqrt{6\xi})^{\frac12(\frac92+\sqrt{6\xi})}
  \bigg(\frac{H_{\rm end}}{m}\bigg)^{\frac12(3+\sqrt{6\xi})}\frac{m^2}{M_{\rm P}}\,.
  \label{Treh3}
  \end{equation}
%Note that this is equal to the upper bound in Eq.~\eqref{Trehbound1}.
%
The condition \mbox{$\Gamma>H_{\rm dom}$} results in the bound
\begin{equation}
  g>\frac{1}{\sqrt{3\pi}}(\sqrt{6\xi})^{\frac32+\sqrt{6\xi}/3}
  \bigg(\frac{H_{\rm end}}{m}\bigg)^{1+\sqrt{6\xi}/3}\frac{m}{M_{\rm P}}\,,
  \label{gbound+}
  \end{equation}
which is the reverse bound to Eq.~\eqref{gbound}.
%Applying this bound in
%Eq.~\eqref{Treh3} we find an upper bound on $T_{\rm reh}$ which is identical
%to Eq.~\eqref{Trehbound2}.

\subsection{Case B: decay before vacuum restoration}
\label{sec:dec}

Now we investigate the behaviour of the condensate when Eq.~(\ref{eq:dec0}) is
satisfied and the condensated decays immediately after reaching $\chi_{\rm min}$.
Since the decay occurs before vacuum restoration we can neglect the bare mass
parameter $m$ in the potential (\ref{eq:pot}).
%Right after the second minimum has emerged the available potential energy for the field one may calculate by simply comparing the potential energies between the two extrema
%\ee{V(0)-V(\chi_{\rm min})\approx \f{(\xi R)^2}{4\lambda}\,.\label{eq:pot2}}
%  we is negligible and we can expand it around the minimum at $\chi^2=-\xi R/\lambda$ in order to obtain
%\ee{V(\chi+\sqrt{-\xi R/\lambda})=-\xi R\chi^2+\sqrt{-\xi\lambda R}\chi^3+\f{\lambda}{4}\chi^4\,,\label{eq:pot2}}
%where we have neglected the vacuum energy piece generated by symmetry breaking.
Thus, the mass of the condensate when it reaches the potenital minimum
$\chi_{\rm min}$ is given by Eq.~\eqref{M} with \mbox{$m=0$},
\mbox{$M=\sqrt{-2\xi R}=2\sqrt{3\xi}\,H$}. When the condensate reaches the
potential miminum, its energy density can be estimated as
\begin{equation}
\rho_\chi^m\approx\frac12 M^2(\chi_{\rm min}^m)^2=\frac{(6\xi)^2}{\lambda}H_m^4\,,
\label{rchi}
\end{equation}
where with `$m$' we denote the moment that the condensate reaches the
potential minimum \mbox{$\chi=\chi_{\rm min}$} and we used that
\mbox{$\chi_{\rm min}^m=\sqrt{\frac{6\xi}{\lambda}}H_m$} (cf.
Eq.~\eqref{eq:min0}).\footnote{We assumed that
\mbox{$\chi_{\rm min}^m>\chi_{\rm end}$}, which is equivalent to 
\mbox{$\sqrt\lambda/2\pi<H_m/H_{\rm end}$}, which is true in view of
Eq.~\eqref{eq:rehcond} and \mbox{$\xi>1$}.} Thus,
\begin{equation}
\left.\frac{\rho_\phi}{\rho_\chi}\right|_m=\frac{3\lambda}{(6\xi)^2}
  \bigg(\frac{M_{\rm P}}{H_m}\bigg)^2\,,
\label{eq:rhpo}
  \end{equation}
where we used \mbox{$\rho_\phi^m=3H_m^2M_{\rm P}^2$}.

%When the generation and decay of the condensate occur very close to the end of inflation we can write the initial energy density as the potential energy of (\ref{eq:pot2}) when the field is at the top of the potential, with $H=H_{\rm end}$ \ee{(\rho_\chi)_{\rm end}=\f{(\xi R)^2}{4\lambda}\bigg\vert_{\rm end}=\f{9\xi^2 H_{\rm end}^4}{\lambda}<{3H_{\rm end}^2}{M_{\rm P}^2}\,,\label{eq:rhpo}}
%where 'end' denotes the end of inflation and the last inequality comes from assuming the field to be a spectator during inflation. We note that if the decay into radiation is not immediate, after inflation (\ref{eq:rhpo}) describes a very quickly diluting fluid $\propto t^{-4}\propto a^{-12}$.

Reheating occurs at the moment when the radiation density overtakes that of the rolling inflaton field, which we can straightforwardly obtain analogously to the previous
section by using Eq.~(\ref{eq:rhpo}) as
\ee{\bigg(\f{a_m}{a_{\rm reh}}\bigg)^2\f{\rho_\phi}{\rho_\chi}\bigg\vert_m=
  \bigg(\f{H_{\rm reh}}{H_m}\bigg)^{2/3}\f{3\lambda M_{\rm P}^2}{(6\xi)^2H_m^2}=1\,,}
leading to the reheating temperature
\ee{T_{\rm reh}=\bigg(\frac{90}{\pi^2g_*}\bigg)^{1/4}
\bigg(\f{6\xi}{\sqrt{3\lambda}}\bigg)^{3/2}\f{H_m^2}{M_{\rm P}}\,.}
Using Eq.~\eqref{eq:rehcond}, the above can be recast as
\ee{\label{Treh4}
  T_{\rm reh}=\bigg(\frac{90}{\pi^2g_*}\bigg)^{1/4}
  \bigg(\f{6\xi}{\sqrt{3\lambda}}\bigg)^{3/2}
  \bigg(\frac{\sqrt\lambda}{2\pi}\bigg)^{\frac{2}{1+\sqrt{6\xi}/3}}
  \f{H_{\rm end}^2}{M_{\rm P}}\,.}

The condensate should not dominate while it rolls down from the central
potential hill. This is because, in such a case, the assumption of a background
undergoing kination domination may not necessarily hold. This means we need
\mbox{$\rho_\chi^m<\rho_\phi^m$}. In view of
Eq.~\eqref{rchi} and also Eq.~\eqref{eq:rehcond} we find the bound
\begin{equation}
  \xi<\frac{\pi}{\sqrt 3}\frac{M_{\rm P}}{H_{\rm end}}
  \bigg(\frac{\sqrt\lambda}{2\pi}\bigg)^{\frac{1}{1+3/\sqrt{6\xi}}}\,.
  \label{lbound}
\end{equation}

Case B is illustrated in Fig. \ref{fig:scal} as the curve that starts out as
orange.
\subsection{Relation to gravitational reheating}
\label{sec:gravr}
Particle production via the generation of the kination condensate as described in this section is inherently a different mechanism to gravitational reheating, where quanta are produced from the  transitional  dynamics  between  the  epoch  of  inflation  and afterwards (e.g. a period of kination) \cite{gravreh}. Gravitational reheating results in general in a bath of  relativistic  particles  and gravitational waves, whose density after its generation dilutes as $a^{-4}.$\footnote{Some particles produced gravitationally may become non-relativistic sometime after inflation, in which case their density scales as $a^{-3}$ and can become dark matter candidates \cite{RefPaper}.}  We can estimate the energy density resulting from gravitational reheating at some time after the end of inflation as \cite{gravreh}
\begin{equation}
\rho_{\rm gr}= q\, g_*^{\rm gr}\frac{\pi^2}{30}\bigg(\frac{H_{\rm end}}{2\pi}\bigg)^4
\bigg(\frac{a_{\rm end}}{a}\bigg)^4\,,
\end{equation}
where $q\lesssim 1$ is an efficiency factor due to the fact that the density of radiation generated by gravitational reheating is of the order of thermal radiation with the Hawkingt emperature for de Sitter space $H_{\rm end}/(2\pi)$ but is not exactly thermal itself. The factor $g_*^{\rm gr}$ includes all relativistic degrees of freedom produced by gravitational reheating, which are fields that are light (meaning with mass less than $H$) but not conformally invariant during inflation. Note that $g_*^{\rm gr}$ includes also the generated gravitational waves.  The gravitational waves unavoidably generated from gravitational reheating can in some cases be boosted by an extended period of kination by such a large amount that obtaining successful Big Bang nucleosynthesis becomes non-trivial bounding the allowed parameter space \cite{Artymowski:2017pua}.

In this work we are interested in parameter ranges for reheating from the kination condensate where the generated energy density is not overwhelmed by the one from gravitational reheating as illustrated by the green curve in Fig.  \ref{fig:scal}.  This leads to the condition
\begin{equation}
\left.\frac{\rho_{\rm gr}}{\rho_\chi}\right|_{\rm reh}< 1
\quad\Leftrightarrow\quad
q\, g_*^{\rm gr}\frac{\pi^2}{30}\bigg(\frac{H_{\rm end}}{2\pi}\bigg)^4
\bigg(\frac{H_{\rm reh}}{H_{\rm end}}\bigg)^{4/3}< 
g_*^{\rm reh}\frac{\pi^2}{30}T_{\rm reh}^4\,,\label{eq:gravr}
\end{equation}
where we have assumed that kination lasts until the moment of reheating. Using that
\mbox{$3M_{\rm P}^2 H_{\rm reh}^2=g_*^{\rm reh}\frac{\pi^2}{30}T_{\rm reh}^4$} 
and after a little algebra, the above suggests
\begin{equation}
T_{\rm reh}> \frac{q^{3/4}}{24\pi^2}\sqrt{\frac{g_*^{\rm gr}}{10}} 
\bigg(\frac{g_*^{\rm gr}}{g_*^{\rm reh}}\bigg)^{1/4}\frac{H_{\rm end}^2}{M_{\rm P}}\,.\label{eq:grbound}
\end{equation}
Assuming \mbox{$q\sim 1$}, \mbox{$g_*^{\rm reh}\approx g_*^{\rm gr}={\cal O}(100)$} and inflation at the energy of grand unification we obtain the bound \mbox{$T_{\rm reh}> 10^{5-6}\,$GeV}.

If the condensate decays {\em after} domination, we have a period after kination and before reheating that the Universe is dominated by the oscillating condensate, whose density scales like matter and \mbox{$a\propto H^{-2/3}$}. Then, it is straightforward to show that the above lower bound on the reheating temperature is further relaxed by the factor \mbox{$\Gamma/H_{\rm dom}$}.

The precise relation of the reheating temperature to the model parameters depends on the specifics of the evolution. Namely, the reheating temperature to be put on the right-hand side in the above is given by
(\ref{Treh1}), (\ref{Treh2}), (\ref{Treh3}) or (\ref{Treh4}).

Qualitatively,  (\ref{eq:gravr})  leads  to  relevant  bounds  mostly  for  case  (A)  where  the energy density of the condensate (potentially) has a lengthy period of $a^{-6}$ scaling,  as is apparent from Fig. \ref{fig:scal}.  Broadly speaking this will only be important for low reheating temperatures (\ref{eq:grbound}), which is also verified in the next section by going through the numerics of example cases.

\subsection{Concrete examples}

Here we will consider a few values for the non-minimal coupling $\xi$ and the
other model parameters (the couplings $\lambda,g$ and the bare mass 
$m$) to demonstrate the efficiency of our reheating mechanism. We consider
inflation at the scale of grand unification, as suggested by the CMB
observations, with \mbox{$H_{\rm end}\sim 10^{-5}\,M_{\rm P}\sim 10^{13}\,$GeV} and
we take \mbox{$g_*=106.75$}, which corresponds to the standard model at high
energies.

\subsubsection{\boldmath $\sqrt{6\xi}=12$ ($\xi=24$)}

%We proceed as before.
The bound in Eq.~\eqref{gbound0} becomes \mbox{$g^2<2$},
which is satisfied for a perturbative coupling. Therefore, the $\chi$-field
condensate decays only after vacuum restoration, that is after the bare mass
$m$ becomes important. Also, the bound in Eq,~\eqref{lbound1} becomes
\begin{equation}
  \frac{\sqrt\lambda}{2\pi}>\bigg(\frac{m}{12H_{\rm end}}\bigg)^5,
\label{l2}
\end{equation}
which turns up to be satisfied for most reasonable values of $m$ and $\lambda$
but not for very large $m$ (see below).
This means that vacuum restoration occurs after the $\chi$-field reaches
$\chi_{\rm min}$. Now, Eq.~\eqref{gbound1} suggests
\begin{equation}
\Gamma<H_{\rm dom}\;\Leftrightarrow\;g<4\sqrt\pi\lambda^{-1/2}\frac{m}{M_{\rm P}}\,.
\label{g3}
\end{equation}
If the above is satisfied, then the reheating temperature is given by
%Eq.~\eqref{T2}.
Eq.~\eqref{Treh2}, which becomes
\begin{equation}
  T_{\rm reh}\sim 0.1\times g\sqrt{mM_{\rm P}}\,.
\label{T2}
\end{equation}
In contrast, if the bound in Eq.~\eqref{g3} is violated, then the reheating
temperature is given by %Eq.~\eqref{T1}.
Eq.~\eqref{Treh1}, which becomes
\begin{equation}
  T_{\rm reh}\sim g^{-1/2}\lambda^{-3/4}\frac{m^2}{M_{\rm P}}\,.
\label{T1}
\end{equation}
To see what the above imply, we select %the same
three indicative values for $m$:

\paragraph{\boldmath $m\sim 1\,$TeV}
\hspace{1cm}\\

\noindent
The bound in Eq.~\eqref{l2} becomes \mbox{$\sqrt\lambda>10^{-54}$}, which is
well satisfied.
%The discussion is the same as with the case
%\mbox{$\sqrt{6\xi}=3$} with
%\mbox{$m=1\,$TeV}, and we obtain a reheating temperature
%\mbox{$T_{\rm reh}\sim 1\,$GeV}, which may be overwhelmed by gravitational
%reheating when the condensate decays before it dominates the Universe.
The bound in Eq.~\eqref{g3} becomes \mbox{$g<10^{-14}\lambda^{-1/2}$}.
If this is satisfied, $T_{\rm reh}$ is given by Eq.~\eqref{T2}, which is maximised
for large $g$, which in turn is maximised for a small $\lambda$. We choose to
consider \mbox{$\lambda\sim 10^{-10}$}. Then the bound on $g$ is
\mbox{$g<10^{-9}$}. Choosing also \mbox{$g\sim 10^{-10}$}, Eq.~\eqref{T2}
suggests that \mbox{$T_{\rm reh}\sim 1\,$GeV}. Suppose now that the bound in
Eq.~\eqref{g3} is violated, such that \mbox{$g>10^{-14}\lambda^{-1/2}$}. This
means that $T_{\rm reh}$ is now given by Eq.~\eqref{T1}. Again $T_{\rm reh}$ is
maximised for small $\lambda$. Choosing again \mbox{$\lambda\sim 10^{-10}$} we
have \mbox{$g>10^{-9}$}. Choosing now \mbox{$g\sim 10^{-8}$}, Eq.~\eqref{T1}
gives again \mbox{$T_{\rm reh}\sim 1\,$GeV}. This is a rather small reheating
temperature, which is overwhelmed by gravitational reheating as discussed in
Sec.~\ref{sec:gravr}, especially when the condensate decays before it dominates
the Universe.

\paragraph{\boldmath $m\sim 10^{10}\,$GeV}
\hspace{1cm}\\

\noindent
The bound in Eq.~\eqref{l2} becomes \mbox{$\sqrt\lambda>10^{-19}$}, which we
expect again to be satisfied.
%The discussion is the same as with the case
%\mbox{$\sqrt{6\xi}=3$} with \mbox{$m=10^{10}\,$GeV}, and we obtain a maximum
%reheating temperature \mbox{$T_{\rm reh}\sim 10^{10}\,$GeV}, which again 
%demonstrates that reheating can be efficient in this model.
The bound in Eq.~\eqref{g3} becomes \mbox{$g<10^{-7}\lambda^{-1/2}$}.
If this is satisfied, $T_{\rm reh}$ is given by Eq.~\eqref{T2}, which is maximised
for large $g$, which in turn is maximised for a small $\lambda$. We choose to
consider again \mbox{$\lambda\sim 10^{-10}$}. Then the bound on $g$ is
\mbox{$g<10^{-2}$}. Choosing also \mbox{$g\sim 10^{-3}$}, Eq.~\eqref{T2}
suggests that \mbox{$T_{\rm reh}\sim 10^{10}\,$GeV}. Suppose now that the bound in
Eq.~\eqref{g3} is violated, such that \mbox{$g>10^{-7}\lambda^{-1/2}$}. This
means that $T_{\rm reh}$ is now given by Eq.~\eqref{T1}. Again $T_{\rm reh}$ is
maximised for small $\lambda$. Choosing again \mbox{$\lambda\sim 10^{-10}$} we
have \mbox{$g>10^{-2}$}. Choosing now \mbox{$g\sim 0.1$}, Eq.~\eqref{T1}
gives again \mbox{$T_{\rm reh}\sim 10^{10}\,$GeV}. This is a sizeable value for
$T_{\rm reh}$, which demonstrates that reheating can be efficient in this model.

\paragraph{\boldmath $m\sim 10^{15}\,$GeV}
\hspace{1cm}\\

\noindent
The bound in Eq.~\eqref{l2} is badly violated. Instead, it is the bound in
Eq.~\eqref{lbound2} that is satisfied. This means that vacuum restoration
occurs before the $\chi$-field reaches $\chi_{\rm min}$.
Now, Eq.~\eqref{gbound} suggests
\begin{equation}
  \Gamma<H_{\rm dom}\;\Leftrightarrow\;g<%\frac{2}{\sqrt\pi}
\bigg(\frac{12\,H_{\rm end}}{m}\bigg)^5\frac{m}{M_{\rm P}}\,.
\label{g4}
\end{equation}
If the above is satisfied then $T_{\rm reh}$ is again given by Eq.~\eqref{T2}.
If not, then $T_{\rm reh}$ is determined by Eq.~\eqref{Treh3}, which becomes
\begin{equation}
T_{\rm reh}\sim 0.1\times g^{-1/2}\bigg(\frac{12H_{\rm end}}{m}\bigg)^{15/2}
\frac{m^2}{M_{\rm P}}.
\label{T4}
\end{equation}
Inputting the value of $m$, the bound in Eq.~\eqref{g4} becomes
\mbox{$g<10^{-8}$}. We choose \mbox{$g\sim 10^{-9}$}. Then Eq.~\eqref{T2} gives
\mbox{$T_{\rm reh}\sim 10^7\,$GeV}.  When the bound in Eq.~\eqref{g4} is violated
and \mbox{$g\geq 10^{-8}$}, $T_{\rm reh}$ is given by Eq.~\eqref{T4}, which becomes
\mbox{$T_{\rm reh}\sim g^{-1/2}10^3\,$GeV}. Taking \mbox{$g\sim 10^{-8}$}, we obtain
again \mbox{$T_{\rm reh}\sim 10^7\,$GeV}, which is the maximum value of
$T_{\rm reh}$ in this case. 

\subsubsection{\boldmath $\sqrt{6\xi}=1000$}

We now consider a large value of $\xi$. The bound in Eq.~\eqref{gbound0} becomes
\mbox{$g<0.1$}, which is possible to violate for a large but permissible
value of $g$. In particular, the Yukawa coupling of the top quark does in fact
violate this bound implying that this mechanism can be effective also when only
the particle content of the Standard Model in considered. Let us consider this
case first. Taking \mbox{$g\sim 1$} we can
satisfy the bound in Eq.~\eqref{eq:dec0}. In this case the condensate decays
immediately when it reaches the minimum of the potential at
\mbox{$\chi_{\rm min}$}, i.e. at $H=H_m$, which is case~B. The reheating
temperature is now given by Eq.~\eqref{Treh4}, which becomes
\begin{equation}
  T_{\rm reh}\sim 10^8\lambda^{-3/4}\frac{H_{\rm end}^2}{M_{\rm P}}\sim
\lambda^{-3/4}\times 10^{16}\,{\rm GeV}\,.
\end{equation}
However, there is a lower bound on $\lambda$ coming from Eq.~\eqref{lbound}.
Putting the numbers in, Eq.~\eqref{lbound} suggests \mbox{$\lambda\gtrsim 1$}.
Thus, \mbox{$T_{\rm reh}\sim 10^{16}\,$GeV}, which is the inflation scale and we
have prompt reheating. This is no surprise as $H_m$ is reached right after the
end of inflation. We also emphasize that the observed value for the Higgs
four-point coupling $\lambda\sim 10^{-2}$ satisfies the bound Eq.~\eqref{lbound}
when the scale of inflation is lower by a few orders of magnitude from our
chosen value of $10^{13}$ GeV, which implies that our mechanism can succesfully
reheat the Universe via the Standard Model Higgs alone and can lead to high
reheating temperatures \mbox{$T_{\rm reh}\sim 10^{12}\,$GeV}.

Consider now \mbox{$g<0.1$}, so that we are back in case~A and the $\chi$-field
condensate decays only after vacuum restoration, that is after the bare mass $m$
becomes important. The bound in Eq.~\eqref{lbound1} is overwhelmingly satisfied
in all cases, so that vacuum restoration occurs after the $\chi$-field reaches
$\chi_{\rm min}$. Eq.~\eqref{gbound1} suggests
\begin{equation}
\Gamma<H_{\rm dom}\;\Leftrightarrow\;g<10^{3/2}\lambda^{-1/2}\frac{m}{M_{\rm P}}\,.
\label{g5}
\end{equation}
If the above is satisfied, then the reheating temperature is given by
Eq.~\eqref{T2}. If the bound in Eq.~\eqref{g5} is violated, then the reheating
temperature is given by Eq.~\eqref{Treh1}, which now becomes
\begin{equation}
T_{\rm reh}\sim 10^2 g^{-1/2}\lambda^{-3/4}\frac{m^2}{M_{\rm P}}\,.
  \label{T5}
\end{equation}
As before, to see what the above imply, we select the same three indicative
values for $m$:

\paragraph{\boldmath $m\sim 1\,$TeV}
\hspace{1cm}\\

\noindent
The bound in Eq.~\eqref{g5} becomes \mbox{$g<10^{-14}\lambda^{-1/2}$}.
If this is satisfied, $T_{\rm reh}$ is given by Eq.~\eqref{T2}, which is maximised
for large $g$, which in turn is maximised for a small $\lambda$. We choose to
consider \mbox{$\lambda\sim 10^{-10}$}. Then the bound on $g$ is
\mbox{$g<10^{-9}$}. Choosing also \mbox{$g\sim 10^{-10}$}, Eq.~\eqref{T2}
suggests that \mbox{$T_{\rm reh}\sim 1\,$GeV} as was always the case when
\mbox{$m\sim 1\,$TeV}. Suppose now that the bound in
Eq.~\eqref{g5} is violated, such that \mbox{$g>10^{-14}\lambda^{-1/2}$}. 
This means that $T_{\rm reh}$ is now given by Eq.~\eqref{T5}. Again $T_{\rm reh}$ is
maximised for small $\lambda$. Choosing \mbox{$\lambda\sim 10^{-10}$} we
have \mbox{$g>10^{-9}$}. Choosing now \mbox{$g\sim 10^{-8}$}, Eq.~\eqref{T5}
gives again \mbox{$T_{\rm reh}\sim 10\,$GeV}.
%which is marginally larger that the cases when $\xi$ was not so large.
These values of $T_{\rm reh}$ are rather
small. Such reheating temperature is overwhelmed by gravitational reheating
(Sec.~\ref{sec:gravr}) especially when the condensate decays before it
dominates the Universe.

\paragraph{\boldmath $m\sim 10^{10}\,$GeV}
\hspace{1cm}\\

\noindent
The bound in Eq.~\eqref{g5} becomes \mbox{$g<10^{-7}\lambda^{-1/2}$}. 
If this is satisfied, $T_{\rm reh}$ is given by Eq.~\eqref{T2}, which is maximised
for large $g$, which in turn is maximised for a small $\lambda$. We choose to
\mbox{$\lambda\sim 10^{-10}$}. Then the bound on $g$ is \mbox{$g<10^{-2}$}. 
Choosing also \mbox{$g\sim 10^{-3}$}, Eq.~\eqref{T2}
suggests that \mbox{$T_{\rm reh}\sim 10^{10}\,$GeV}. Suppose now that the bound in
Eq.~\eqref{g5} is violated, such that \mbox{$g>10^{-7}\lambda^{-1/2}$}. This
means that $T_{\rm reh}$ is now given by Eq.~\eqref{T5}. Again $T_{\rm reh}$ is
maximised for small $\lambda$. Choosing again \mbox{$\lambda\sim 10^{-10}$} we
have \mbox{$g>10^{-2}$}. Selecting \mbox{$g\sim 0.1$}, Eq.~\eqref{T5}
gives \mbox{$T_{\rm reh}\sim 10^{12}\,$GeV}. Note that the value
\mbox{$g\sim 0.1$} saturates the bound in Eq.~\eqref{gbound0}.

\paragraph{\boldmath $m\sim 10^{15}\,$GeV}
\hspace{1cm}\\

\noindent
In contrast to the small $\xi$ case, when $\xi$ is large the bound in
Eq.~\eqref{lbound1} is satisfied even for large values of $m$. Thus, the
condensate decays only after vacuum restoration.
The bound in Eq.~\eqref{g5} becomes now \mbox{$g<10^{-2}\lambda^{-1/2}$}. 
One does not need to consider very low values of $\lambda$. Choosing
\mbox{$\lambda\sim 10^{-2}$} we obtain \mbox{$g<0.1$}. This bound must be
satisfied since a larger $g$ would violate the bound in Eq.~\eqref{gbound0}.
Then, the reheating temperature is given by Eq.~\eqref{T2}, which suggests
\mbox{$T_{\rm reh}\sim g\times 10^{16}\,$GeV}. Thus, saturating the upper bound on
$g$ we obtain \mbox{$T_{\rm reh}\sim 10^{15}\,$GeV}.

Our results do not change dramatically when employing much bigger values
of $\lambda$. For example, consider the case of \mbox{$\xi=24$} and 
\mbox{$m\sim 10^{10}\,$GeV}. The bound on $g$ is given by Eq.~\eqref{g3},
which reads \mbox{$g<10^{-7}\lambda^{-1/2}$}. Suppose that we choose
\mbox{$\lambda\sim 10^{-4}$}. Then following the same procedure as discussed
above we find \mbox{$T_{\rm reh}\sim 10^7\,$GeV} for \mbox{$g\sim 10^{-5}$}. If we
choose \mbox{$\lambda\sim 10^{-2}$}, we obtain \mbox{$T_{\rm reh}\sim 10^6\,$GeV}
for \mbox{$g\sim 10^{-7}$}. It is more realistic to consider relatively large
values of $\lambda$, which is the self-coupling of the $\chi$ scalar field, than
$g$, which is a Yukawa coupling of $\chi$ to its fermionic decay products. We
see that we may consider even $\lambda\sim 0.01$ at the expense of a few
orders of magnitude on the value of $T_{\rm reh}$, which is still decently large.

\section{Conclusions}

In this paper we studied a new mechanism to reheat the Universe during
kination, which corresponds to an epoch dominated by the kinetic density
of the inflaton field. Kination follows non-oscillatory inflation in which the
inflaton field does not oscillate around its VEV but rolls down over large
distances in field space. We have introduced a spectator field, non-minimally
coupled to gravity, which can decay into the thermal bath of the hot big bang
and terminate kination. Our spectator field is not coupled to the inflaton.

During inflation, the non-minimal coupling (assumed positive) renders the
spectator field heavy and confines it at the origin. After the end of inflation
and the onset of kination, however, the origin becomes a local maximum and the
field is displaced towards the temporary minimum generated by the self-coupling
of the field, which stabilises the scalar potential. Eventually, the bare mass
of the field overwhelms the non-minimal term, the origin becomes again the
minimum of the scalar potential (the VEV) and the spectator field coherently
oscillates around its VEV until its decay into radiation, which reheats the
Universe. We have investigated all possibilities, depending on the values of
the bare mass and the couplings of the spectator field and showed that, with
natural values of the model parameters, we can achieve successful reheating.
For example, if the bare mass is around \mbox{$m\sim 10^{10}\,$GeV}, the
self-coupling is taken to be \mbox{$\lambda\sim 0.01$} and the Yukawa coupling
to light fermions is \mbox{$g\sim 10^{-7}$} we obtain
\mbox{$T_{\rm reh}\sim 10^6\,$GeV}. For large values of the bare mass one may
even achieve prompt reheating, which occurs right after the end of inflation.
For small values of $m$ with small $g$, however, the efficiency of reheating is
not great.
For example, when \mbox{$m\sim 1\,$TeV}, the typical value of the reheating
temperature is \mbox{$T_{\rm reh}\sim 1\,$GeV}, which is overwhelmed by
gravitational reheating if our spectator field decays before dominating the
Universe content.

In the context of Standard Model physics and other models with large Yukawa
couplings, reheating can also be quite efficient with high reheating
temperatures. As an example, we showed that if the Standard Model Higgs is
non-minimally coupled to gravity with a non-minimal coupling
$\sqrt{6\xi}\sim 10^3$ and inflation ends at a scale $H_{\rm end}\sim 10^{11}\,$GeV
the Universe reheats with \mbox{$T_{\rm reh}\sim 10^{12}\,$GeV}.
 
 We emphasise that our model is highly predictive and does not depend on initial 
conditions. In a variant of our model, it is easy to show that our
spectator field may generate the observed dark matter if it decays into heavy 
fermions or not at all, provided that the Universe is reheated by some other 
mechanism. A similar set-up for gravitational dark matter production was
discovered in Ref.~\cite{Markkanen:2015xuw} and will be investigated in detail
in Ref.~\cite{inprep}.  

When our paper appeared in arXiv another relevant paper where a similar set-up
was studied \cite{Nakama:2018gll} appeared simultaneously. In
Ref.~\cite{Nakama:2018gll} however, the scalar field  considered is the Higgs
field, with only the interactions stipulated by the Standard Model. In our case,
we consider the condensate dynamics in an extended framework encompassing also
cases where the condensate does not decay immediately after inflation by
allowing the mass, the four-point coupling as well as the Yukawa coupling to
vary.

%%%%%%%%%%%%%%%%%%%%%%%%%%%%%%%%%%%%%%%%%%%%%%%%%%%%%
%\acknowledgments
%%%%%%%%%%%%%%%%%%%%%%%%%%%%%%%%%%%%%%%%%%%%%%%%%%%%%
KD  is  supported  (in  part)  by  the Lancaster-Manchester-Sheffield
Consortium for Fundamental Physics under STFC grant ST/L000520/1.
TM is supported by the STFC grant ST/P000762/1.   

%%%%%%%%%%%%%%%%%%%%%%%%%%%%%%%%%%%%%%%%%%%%%%%%%%%%%
%%%%%%%%%%%%%%%%%%%%%%%%%%%%%%%%%%%%%%%%%%%%%%%%%%%%%

\end{document}